\documentclass[%
reprint,
superscriptaddress,
showpacs,
amsmath,amssymb,
prc,
floatfix, ]%
{revtex4-1}

\usepackage{mathrsfs}

\usepackage{color}

\usepackage{graphicx}
\usepackage{dcolumn}
\usepackage{bm}
\usepackage[dvipdfm,bookmarks=true,colorlinks,%
            citecolor=blue,linkcolor=blue,hypertex, %
            breaklinks=true]{hyperref}

\begin{document}



\title{Effects of proton angular momentum alignment on
       the two-shears-like mechanism in $^{101}$Pd}

\author{Zhen-Hua Zhang}
 \email{zhzhang@ncepu.edu.cn}
 \affiliation{Mathematics and Physics Department,
              North China Electric Power University, Beijing 102206, China}
\date{\today}

\begin{abstract}

The recently observed possible antimagnetic rotation band
in $^{101}$Pd is investigated by the cranked shell model with
pairing correlations treated by a particle-number conserving method,
in which the blocking effects are taken into account exactly.
The experimental moments of inertia and reduced $B(E2)$ transition probabilities
and their variations with the rotational frequency $\omega$ are well reproduced.
By analyzing the $\omega$-dependence of the occupation probability of each
cranked Nilsson orbital near the Fermi surface and the contributions of valence
orbitals in each major shell to the total angular momentum alignment,
the upbending mechanism of $\nu h_{11/2}$ in $^{101}$Pd is understood clearly.
The proton angular momentum alignment and its influence on the two-shears-like
mechanism are also discussed.

\end{abstract}


\maketitle


\section{\label{Sec:Introduction}Introduction}

Magnetic rotation (MR)~\cite{Frauendorf1994_Proceedings}
and antimagnetic rotation (AMR)~\cite{Frauendorf1996_Proceedings} are interesting exotic
rotational phenomena observed in weakly deformed or near spherical nuclei~\cite{Frauendorf2001_RMP73-463}.
In the MR bands, the energy and angular
momentum are generated due to the so-called ``shears mechanism'',
i.e., the alignments of the high-$j$ proton and neutron angular momenta.
This  type of rotational bands have been discovered
experimentally in $A \sim 60, 80, 110, 130$, and 190 mass regions~\cite{Clark2000_ARNPS50-1,
Frauendorf2001_RMP73-463, Hubel2005_PPNP54-1, Meng2013_FrontiersofPhysics8-55}.
In analogy to the antiferromagnetism in condensed matter physics, a similar
phenomenon known as ``antimagnetic rotation'' is predicted in nuclei
by Frauendorf~\cite{Frauendorf2001_RMP73-463, Frauendorf1996_Proceedings}.
In AMR bands, the energy and angular momentum are increased by
the so-called ``two-shears-like mechanism'',
i.e., by simultaneously closing of the two
valence protons (neutrons) toward the neutron (proton) angular momentum vector.
The physical reason behind the establishment of such rotational bands built on near-spherical nuclei
is the violation of the rotational symmetry by the nucleon currents.

AMR is expected to be observed in the same mass regions
as MR~\cite{Frauendorf2001_RMP73-463}.
Since AMR was proposed~\cite{Frauendorf2001_RMP73-463},
it has been investigated both from experimental and theoretical aspects.
Up to now, experimental evidence of AMR has been reported in Cd isotopes
including $^{105}$Cd~\cite{Choudhury2010_PRC82-061308R},
$^{106}$Cd~\cite{Simons2003_PRL91-162501},
$^{108}$Cd~\cite{Simons2005_PRC72-024318, Datta2005_PRC71-041305R},
$^{110}$Cd~\cite{Roy2011_PLB694-322}, and
${}^{107}$Cd~\cite{Choudhury2013_PRC87-034304}.
In addition, the occurrence of this phenomenon still needs
further investigation by lifetime measurements in
$^{109}$Cd~\cite{Chiara2000_PRC61-034318},
$^{100}$Pd~\cite{Zhu2001_PRC64-041302R},
$^{144}$Dy~\cite{Sugawara2009_PRC79-064321},
and
$^{112}$In~\cite{Li2012_PRC86-057305}.

Theoretically, AMR has been discussed by simple
geometry in the classical particle rotor model~
\cite{Clark2000_ARNPS50-1},
and the tilted axis cranking (TAC) model~
\cite{Frauendorf2000_NPA677-115, Peng2008_PRC78-024313,
Zhao2011_PLB699-181}.
Based on the TAC model, many applications have been carried out
in the framework of microscopic-macroscopic model~
\cite{Zhu2001_PRC64-041302R, Simons2003_PRL91-162501, Simons2005_PRC72-024318},
pairing plus quadrupole model~
\cite{Chiara2000_PRC61-034318, Frauendorf2001_RMP73-463},
and the covariant density functional theory (CDFT)
~\cite{Zhao2011_PRL107-122501, Zhao2012_PRC85-054310, Liu2012_SSPMA55-2420,
Peng2015_PRC91-044329}.
In particular, by using the point-coupling density functional~\cite{Zhao2010_PRC82-054319},
TAC-CDFT has also been applied successfully in describing many other phenomena such as
magnetic and chiral rotation~\cite{Meng2016_PS91-053008},
nuclear rod shape~\cite{Zhao2015_PRL115-022501}, etc.
Very recently, the TAC-CDFT has also been extended to including pairing correlations in Ref.~\cite{Zhao2015_PRC92-034319}.
We note that pairing correlations, in most of the existing studies,
are either neglected or treated by the Bogoliubov formalism,
where the particle number is not conserved.
The violation of the particle number may raise serious problems~\cite{Zeng1983_NPA405-1, Molique1997_PRC56-1795}.
Actually, all cranked Hartree-Fock-Bogolyubov (HFB) calculations
show that a pairing collapsing occurs for angular
momentum $I$ greater than a critical value $I_c$~\cite{Mottelson1960_PRL5-511}.
The remedy in terms of the particle-number projection or the
Lipkin-Nogami method can restore this broken symmetry.
Previous investigations show that, after performing the particle-number projection,
the description of the rotational properties can be improved considerably
comparing with the HFB cranking calculations~\cite{Dudek1988_PRC38-940}.
However, they complicate the algorithms considerably, yet without improving the
description of the higher-excited part
of the spectrum of the pairing Hamiltonian~\cite{Molique1997_PRC56-1795}.

In the present work, the cranked shell model (CSM) with
pairing correlations treated by a particle-number conserving (PNC)
method~\cite{Zeng1983_NPA405-1, Zeng1994_PRC50-1388} is used
to investigate the possible AMR band $\nu h_{11/2}$ in
$^{101}$Pd~\cite{Sugawara2012_PRC86-034326, Sugawara2015_PRC92-024309}.
The PNC-CSM has already been used to investigate the AMR bands
in $^{105}$Cd and $^{106}$Cd~\cite{Zhang2013_PRC87-054314}. However,
quite recently, lifetime measurements have been performed for the
$\nu h_{11/2}$ band in $^{101}$Pd~\cite{Sugawara2015_PRC92-024309},
which confirm the previous assumption of the
AMR nature of this band~\cite{Sugawara2012_PRC86-034326}.
Note that the $^{101}$Pd nucleus has four proton $g_{9/2}$ holes,
so the two-shears-like mechanism for this AMR
band may be different from that of the Cd isotopes.
Furthermore, the pairing interaction should be more prominent in
 $^{101}$Pd due to two more proton holes than Cd isotopes.
Therefore, it is interesting to investigate effects of pairing and
the two-shears-like mechanism in this nucleus.

In contrary to the conventional Bardeen-Cooper-Schrieffer (BCS) or
Bogoliubov approaches, the Hamiltonian is solved directly
in a truncated Fock-space in the PNC method~\cite{Wu1989_PRC39-666}.
Therefore, the particle-number is conserved and the Pauli blocking effects
are taken into account exactly.
The PNC scheme has also been used both in relativistic
and nonrelativistic mean field models~\cite{Meng2006_FPC1-38, Pillet2002_NPA697-141} and
the total-Routhian-surface method with the
Woods-Saxon potential~\cite{Fu2013_PRC87-044319, Fu2013_SCPMA56-1423}.
Very recently, the particle-number conserving method based on the
cranking Skyrme-Hartree-Fock model has been developed~\cite{Liang2015_PRC92-064325}.
The PNC-CSM has also been employed successfully for describing various
phenomenon concerning on the rotating nuclei, e.g.,
the odd-even differences in moments of inertia (MOI's)~\cite{Zeng1994_PRC50-746},
the identical bands~\cite{Liu2002_PRC66-024320, He2005_EPJA23-217},
the nuclear pairing phase transition~\cite{Wu2011_PRC83-034323},
the rotational bands and high-$K$ isomers in the rare-earth~\cite{Liu2004_NPA735-77,
Zhang2009_NPA816-19, Zhang2009_PRC80-034313, Zhang2016_NPA949-22},
the actinide nuclei~\cite{He2009_NPA817-45,
Zhang2011_PRC83-011304R, Zhang2012_PRC85-014324, Zhang2013_PRC87-054308}, etc.

This paper is organized as follows.
A brief introduction to the PNC treatment of pairing correlations within
the CSM is presented in Sec.~\ref{Sec:PNC-CSM}.
The results and discussion are given in Sec.~\ref{Sec:Results}.
A brief summary is given in Sec.~\ref{Sec:Summary}.

\section{\label{Sec:PNC-CSM}Theoretical framework}

The cranked shell model Hamiltonian of an axially symmetric
nucleus in the rotating frame can be written as
\begin{eqnarray}
 H_\mathrm{CSM}
 & = &
 H_0 + H_\mathrm{P}
 = H_{\rm Nil}-\omega J_x + H_\mathrm{P}
 \ ,
 \label{eq:H_CSM}
\end{eqnarray}
where $H_{\rm Nil}$ is the Nilsson Hamiltonian, $-\omega J_x$ is the
Coriolis interaction with the cranking frequency $\omega$ about the
$x$ axis (perpendicular to the nuclear symmetry $z$ axis).
$H_{\rm P}$ is the pairing interaction,
\begin{eqnarray}
 H_{\rm P}
 & = &
  -G \sum_{\xi\eta} a^\dag_{\xi} a^\dag_{\bar{\xi}}
                        a_{\bar{\eta}} a_{\eta}
  \ ,
\end{eqnarray}
where $\bar{\xi}$ ($\bar{\eta}$) labels the time-reversed state of a
Nilsson state $\xi$ ($\eta$),
and $G$ is the effective strength of the monopole pairing interaction.

Instead of the usual single-particle level truncation in conventional
shell-model calculations, a cranked many-particle configuration
(CMPC) truncation (Fock space truncation) is adopted which is crucial
to make the PNC calculations for low-lying excited states both
workable and sufficiently accurate~\cite{Wu1989_PRC39-666, Molique1997_PRC56-1795}.
Usually a dimension of 1000 should be enough for the calculations of the heavy nuclei.
An eigenstate of $H_\mathrm{CSM}$ can be written as
\begin{equation}
 |\Psi\rangle = \sum_{i} C_i \left| i \right\rangle
 \ ,
 \qquad (C_i \; \textrm{real}),
\end{equation}
where $| i \rangle$ is a CMPC (an eigenstate of the one-body operator $H_0$).
The expectation value of a one-body operator
$\mathcal {O} = \sum_{k=1}^N \mathscr{O}(k)$ is  thus written as
\begin{equation}
 \left\langle \Psi | \mathcal {O} | \Psi \right\rangle
 =\sum_i C_i^2 \left\langle i | \mathcal {O} | i \right\rangle
 +2\sum_{i<j} C_i C_j \left\langle i | \mathcal {O} | j \right\rangle \ .
\end{equation}
As $\mathcal {O}$ is a one-body operator,
the matrix element $\langle i | \mathcal {O} | j \rangle$ for $i\neq j$
is nonzero only when $|i\rangle$ and $|j\rangle$ differ by
one particle occupation~\cite{Zeng1994_PRC50-1388}.
After a certain permutation of creation operators,
$|i\rangle$ and $|j\rangle$ can be recast into
\begin{equation}
 | i \rangle = (-1)^{M_{i\mu}} | \mu \cdots \rangle \ , \qquad
 | j \rangle = (-1)^{M_{j\nu}} | \nu \cdots \rangle \ ,
\end{equation}
where $\mu$ and $\nu$ denote two different single-particle states,
the ellipsis $\cdots$ stands for the same particle occupation,
and $(-1)^{M_{i\mu}}=\pm1$, $(-1)^{M_{j\nu}}=\pm1$ according to
whether the permutation is even or odd.
Therefore, the  expectation value of $\mathcal {O}$
can be separated into the diagonal $\sum_{\mu} \mathscr{O}(\mu)$
and the off-diagonal $2\sum_{\mu<\nu} \mathscr{O}(\mu\nu)$ parts
\begin{eqnarray}
 \mathcal {O}
  &=& \left\langle \Psi | \mathcal {O} | \Psi \right\rangle =
  \left( \sum_{\mu} \mathscr{O}(\mu) + 2\sum_{\mu<\nu} \mathscr{O}(\mu\nu) \right) \ , \label{eq:j1}\\
 \mathscr{O}(\mu)
 &=& \langle \mu | \mathscr{O} | \mu \rangle n_{\mu}  \ , \label{eq:j1d} \\
 \mathscr{O}(\mu\nu)
 &=&\langle \mu | \mathscr{O} | \nu \rangle
  \sum_{i<j} (-1)^{M_{i\mu}+M_{j\nu}} C_{i} C_{j} \ ,
  \label{eq:j1od}
\end{eqnarray}
where
\begin{equation}
n_{\mu}=\sum_{i}|C_{i}|^{2}P_{i\mu}
\end{equation}
is the occupation probability of the cranked Nilsson orbital $|\mu\rangle$
and $P_{i\mu}=1$ (0) if $|\mu\rangle$ is occupied (empty) in $|i\rangle$.

The kinematic moment of inertia $J^{(1)}$ of $|\Psi\rangle$ can be written as
\begin{eqnarray}
 J^{(1)} = \frac{1}{\omega} \left\langle \Psi | J_x | \Psi \right\rangle \ .
\end{eqnarray}
The $B(E2)$ transition probabilities can be derived
in the semiclassical approximation as
\begin{equation}
B(E2) = \frac{3}{8}
{\left\langle \Psi | Q_{20}^{\rm p} | \Psi \right\rangle}^2 \ ,
\end{equation}
where $Q_{20}^{\rm p}$ corresponds to the quadrupole moments of protons and
\begin{equation}
Q_{20} = \sqrt{\frac{5}{16\pi}} (3z^2-r^2) = r^2 Y_{20} \ .
\end{equation}

\section{\label{Sec:Results} Results and discussion}

The Nilsson parameters ($\kappa$ and $\mu$) for $^{101}$Pd are taken
from the traditional values~\cite{Nilsson1969_NPA131-1}.
The deformation parameters $\varepsilon_2 = 0.125$ and $\varepsilon_4 = -0.02$
are taken from Ref.~\cite{Moeller1995_ADNDT59-185}.
The valence single-particle space in this work is constructed
in the major shells from $N=0$ to $N=5$ both for protons and neutrons,
so there is no effective charge involved in the calculation of the $B(E2)$ values.
In principle, the effective pairing strengths can
be determined by the odd-even differences in nuclear binding energies,
and are connected with the dimension of the truncated CMPC space.
The CMPC truncation energies are
about 1.0$\hbar\omega_0$ for protons and 0.9$\hbar\omega_0$ for neutrons, respectively.
For $^{101}$Pd, $\hbar\omega_{\rm 0p}=8.542$~MeV
for protons and $\hbar\omega_{\rm 0n}=9.065$~MeV for neutrons~\cite{Nilsson1969_NPA131-1}.
The dimensions of the CMPC space are about 1000 both for protons and neutrons.
The corresponding effective pairing strengths used in this work are
$G_{\rm p}$ = 0.6~MeV and $G_{\rm n}$ = 0.6~MeV.
A larger CMPC space with renormalized pairing strengths gives essentially the same results.
In addition, the stability of the PNC-CSM results against the change
of the dimension of the CMPC space has been investigated in
Refs.~\cite{Zeng1994_PRC50-1388, Zhang2012_PRC85-014324}.
In present calculations, almost all the important CMPC's
(with the corresponding weights larger than $0.1\%$) are taken into account,
so the solutions to the low-lying excited states are accurate enough.

\begin{figure}[htbp]
\includegraphics[width=1.0\columnwidth]{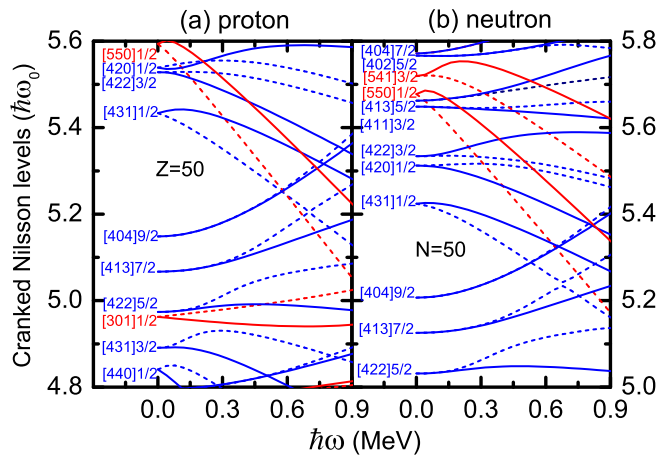}
\caption{\label{fig1:Nil}(Color online)
The cranked Nilsson levels near the Fermi surface of $^{101}$Pd
for (a) protons and (b) neutrons.
The positive (negative) parity levels are denoted by blue (red) lines.
The signature $\alpha=+1/2$ ($\alpha=-1/2$) levels
are denoted by solid (dotted) lines.
The Nilsson parameters ($\kappa$ and $\mu$) are taken from
the traditional values~\cite{Nilsson1969_NPA131-1}.
The deformation parameters $\varepsilon_2 = 0.125$ and $\varepsilon_4 = -0.02$
are taken from Ref.~\cite{Moeller1995_ADNDT59-185}.}
\end{figure}

The cranked Nilsson levels near the Fermi surface of $^{101}$Pd are shown
in Fig.~\ref{fig1:Nil}(a) for protons and (b) for neutrons.
The positive (negative) parity levels are denoted by blue (red) lines.
The signature $\alpha=+1/2$ ($\alpha=-1/2$) levels
are denoted by solid (dotted) lines.
It can be seen from Fig.~\ref{fig1:Nil}(a) that the four proton
holes in $^{101}$Pd are $\pi 9/2^+[404]$ ($g_{9/2}$) and $\pi 7/2^+[413]$ $(g_{9/2})$.
The data show that the possible AMR band in $^{101}$Pd
is the lowest lying negative parity band,
which is assigned as $\nu h_{11/2}$~\cite{Sugawara2012_PRC86-034326, Sugawara2015_PRC92-024309}.
It can be seen from Fig.~\ref{fig1:Nil}(b) that, in the present calculation,
the lowest lying negative parity state in $^{101}$Pd at cranking frequency
$\hbar\omega = 0$~MeV is $ \nu 1/2^-[550]$ $(h_{11/2})$.
Therefore, in the following investigation, adiabatic calculations
for the $\nu 1/2^-[550]$ band in $^{101}$Pd will be performed and the
level crossings in protons and neutrons appear automatically
with increasing rotational frequency.

\begin{figure}[htbp]
\includegraphics[width=0.9\columnwidth]{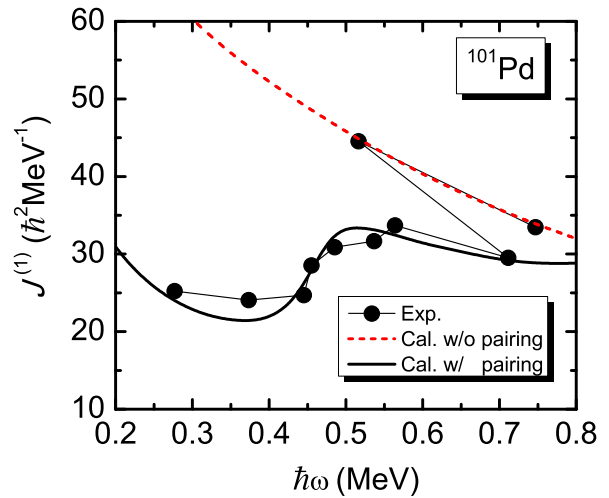}
\caption{\label{fig2:MOI} (Color online)
The experimental (solid circles) and calculated kinematic MOI's
$J^{(1)}$ with (black solid line) and without (red dashed line)
pairing correlations for $\nu 1/2^-[550]$ in $^{101}$Pd.
The data are taken from Ref.~\cite{Sugawara2015_PRC92-024309}.
}
\end{figure}

Figure~\ref{fig2:MOI} shows the experimental (solid circles) and
calculated kinematic MOI's $J^{(1)}$ with (black solid line) and without
(red dashed line) pairing correlations for $\nu 1/2^-[550]$ in $^{101}$Pd.
The data are taken from Ref.~\cite{Sugawara2015_PRC92-024309}.
The pairing correlations are very important in reproducing
the experimental MOI's, especially for the rotational frequency region
before the sharp backbending at $\hbar\omega\sim 0.65$~MeV.
It can be seen that the MOI's of $^{101}$Pd are
overestimated when the pairing is switched off,
while they are well reproduced after considering
the pairing correlations except the last two data,
which are consistent with the calculated results without pairing.
This may indicate that in the very high-spin region, the effect of
pairing correlations on the reduction of MOI's is negligible.
In Refs.~\cite{Sugawara2012_PRC86-034326, Sugawara2015_PRC92-024309},
the upbending around $\hbar\omega \sim 0.45$~MeV is interpreted as
alignments of two $g_{7/2}$ quasi-neutrons.
In the following, the upbending mechanism in this band will be investigated.
It should be noted that the rotational properties of this band are quite
different from those observed in $^{105}$Cd and
$^{106}$Cd~\cite{Zhang2013_PRC87-054314},
in which the AMR happens in the high-spin region after the first upbending
and the MOI's nearly keep constant with increasing rotational frequency.

\begin{figure}[htbp]
\includegraphics[width=0.9\columnwidth]{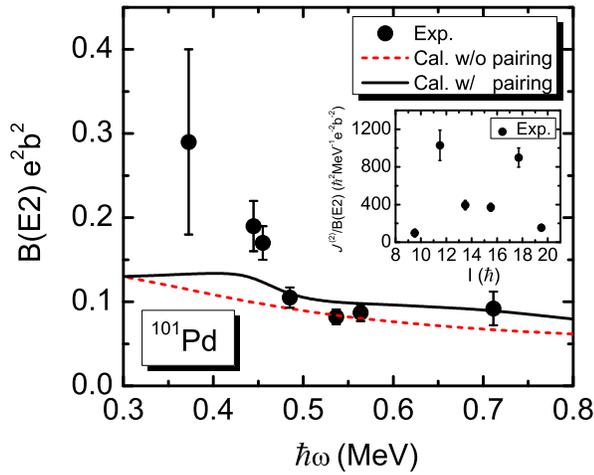}
\caption{\label{fig3:be2}(Color online)
The experimental (solid circles) and calculated $B(E2)$ values
with (black solid line) and without (red dashed line)
pairing correlations for $\nu 1/2^-[550]$ in $^{101}$Pd.
The inset shows the experimental $J^{(2)}/B(E2)$ values.
The data are taken from Ref.~\cite{Sugawara2015_PRC92-024309}.
}
\end{figure}

Figure~\ref{fig3:be2} shows the experimental (solid circles) and calculated $B(E2)$ values
with (black solid line) and without (red dashed line)
pairing correlations for $\nu 1/2^-[550]$ in $^{101}$Pd.
The data are taken from Ref.~\cite{Sugawara2015_PRC92-024309}.
It can be seen that the $B(E2)$ values at higher rotational frequency
$\hbar\omega > 0.45$~MeV can be reproduced well no matter the
pairing is considered or not.
However, the description of the quickly drop of $B(E2)$ values around
$\hbar\omega \sim 0.45$~MeV can be improved by taking the pairing correlations into account,
even the calculated results still deviate a little from the data.
It is difficult to describe the $B(E2)$ behavior with a frozen deformation parameter
in the present calculations.
This may be due to the deformation change with the rotational frequency for $^{101}$Pd.
It should be noted that the quickly drop of $B(E2)$ values around
$\hbar\omega = 0.4$ to 0.5 MeV is just corresponding to the upbending region in the MOI's.
Therefore, it is interesting to know how the angular momentum alignments
affect the two-shears-like mechanism.
Since it is well known that with the two proton angular momentum vectors closing,
the $B(E2)$ values will be decreased.
The inset shows the experimental $J^{(2)}/B(E2)$ values.
It has been shown in Ref.~\cite{Sugawara2015_PRC92-024309} that,
the large $J^{(2)}/B(E2)$ values indicate the AMR nature of this band.

\begin{figure}[htbp]
\includegraphics[width=0.9\columnwidth]{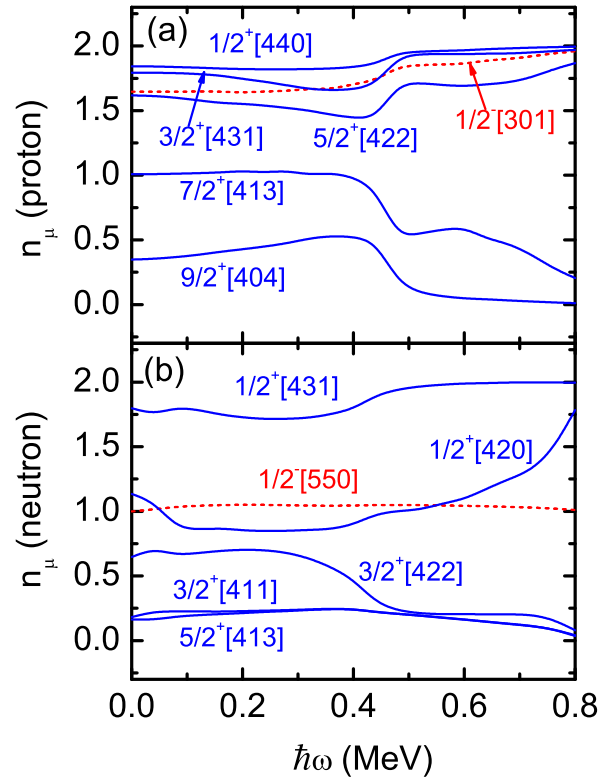}
\caption{\label{fig4:occup}(Color online)
Occupation probability $n_\mu$ of each orbital
$\mu$ (including both $\alpha=\pm1/2$) near the Fermi surface
of $\nu 1/2^-[550]$ in $^{101}$Pd.
The top and bottom rows are for protons and neutrons, respectively.
The positive (negative) parity levels are denoted by blue solid (red dotted) lines.
The Nilsson levels far above the Fermi surface
($n_{\mu}\sim0$) and far below ($n_{\mu}\sim2$) are not shown.
}
\end{figure}

One of the advantages of the PNC method is that the total particle
number $N = \sum_{\mu}n_\mu$ is exactly conserved,
whereas the occupation probability $n_\mu$ for each orbital varies
with rotational frequency.
By examining the $\omega$-dependence of the orbitals close to the Fermi surface,
one can learn more about how the Nilsson levels evolve with rotation and
get some insights on the upbending mechanism.
Figure~\ref{fig4:occup} shows the occupation probability $n_\mu$ of each orbital
$\mu$ (including both $\alpha=\pm1/2$) near the Fermi surface for protons (upper panel)
and neutrons (lower panel), respectively.
The positive (negative) parity levels are denoted by blue solid (red dotted) lines.
The Nilsson levels far above the Fermi surface
($n_{\mu}\sim0$) and far below ($n_{\mu}\sim2$) are not shown.
It can be seen from Fig.~\ref{fig4:occup}(a) that at
the rotational frequency $\hbar\omega \sim 0.45$ MeV, occupation probabilities for
the orbitals $\pi 9/2^+[404]$ and $\pi 7/2^+[413]$ drop down quickly
from 1.0 to about 0.5 and from 0.5 to nearly zero, respectively,
while the occupation probabilities of some other orbitals e.g.,
$\pi 5/2^+[422]$ ($g_{9/2}$) and $\pi 3/2^+[431]$ ($g_{9/2}$), slightly increase.
This can be understood from the cranked Nilsson orbitals in Fig.~\ref{fig1:Nil}(a).
The proton orbitals $\pi 9/2^+[404]$ and $\pi 7/2^+[413]$ are above
the Fermi surface at $\hbar\omega=0$.
Due to the pairing correlations, these two orbitals are partly occupied.
With increasing cranking frequency, these two orbitals leave
farther above the Fermi surface, so the occupation probabilities of
these two orbitals become smaller with increasing $\hbar\omega$.
Meanwhile, the occupation probabilities of those orbitals which approach near to
the Fermi surface become larger with increasing  $\hbar\omega$.
The situation is similar in Fig.~\ref{fig4:occup}(b).
The occupation probability of $\nu 3/2^+[422]$ ($2d_{5/2}$)
decreases slowly from 0.5 to nearly 0.2 with the increasing frequency
$\hbar\omega$ from about $0.3$~MeV to $0.5$~MeV,
while the occupation probabilities of $\nu 1/2^+[420]$ ($1g_{7/2}$) and
$\nu 1/2^+[431]$ ($2d_{5/2}$) increase gradually with $\hbar\omega$.
Therefore, the contributions to the upbending at $\hbar\omega \sim 0.45$
for $\nu 1/2^-[550]$ in $^{101}$Pd may come from the
rearrangement of proton occupations in $g_{9/2}$ orbitals and the
alignment of neutrons in $1g_{7/2}$ and $2d_{5/2}$ orbitals.
Note that in PNC-CSM calculations, the proton configuration of the
AMR bands in $^{105}$Cd and $^{106}$Cd is nearly one pair of
pure proton $g_{9/2}$ holes (occupation probabilities are close
to zero)~\cite{Zhang2013_PRC87-054314}.
While in $^{101}$Pd, due to stronger pairing correlations,
the two pair of proton $g_{9/2}$ holes are partly occupied and the
occupation probabilities are rearranged with increasing rotational frequency,
which may indicate a new picture of AMR.

\begin{figure}[htbp]
\includegraphics[width=0.9\columnwidth]{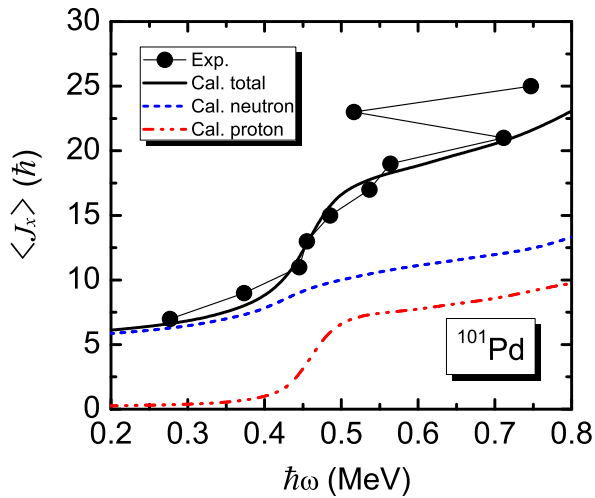}
\caption{\label{fig5:jx}(Color online)
The experimental (solid circles) and calculated (black solid line)
angular momentum alignment $\langle J_x\rangle$
for $\nu 1/2^-[550]$ in $^{101}$Pd.
The contributions of neutrons and protons to $\langle J_x\rangle$
calculated from PNC-CSM are denoted by blue dashed line and red dotted dash line,
respectively.
}
\end{figure}

To analyze the upbending mechanism for $\nu 1/2^-[550]$ in $^{101}$Pd,
the experimental (solid circles) and calculated (black solid line)
angular momentum alignment $\langle J_x\rangle$ are shown in Fig.~\ref{fig5:jx}.
The contributions of neutrons and protons to $\langle J_x\rangle$
calculated from PNC-CSM are denoted by blue dashed line and red dotted dash line,
respectively.
It can be seen that at the upbending region ($\hbar\omega \sim 0.45$~MeV),
the contributions from protons to the total angular momentum alignment $\langle J_x\rangle$
are increased more than that of the neutrons,
which indicates that this upbending mainly comes from the contribution of the protons.
The present results are different from those obtained in
Refs~\cite{Sugawara2012_PRC86-034326, Sugawara2015_PRC92-024309},
where the increase of the angular momentum alignment
is assumed to be from the alignment of one $\nu g_{7/2}$ neutron pair.
In the present PNC-CSM calculation, the contributions from neutrons are much
less considerable than those from protons.

\begin{figure}[htbp]
\includegraphics[width=0.9\columnwidth]{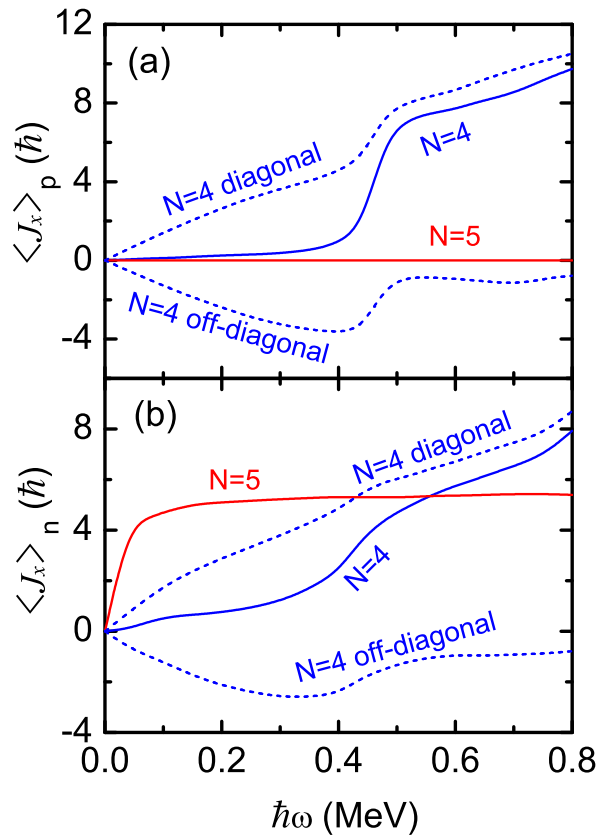}
\caption{\label{fig6:jxshell}(Color online)
Contributions of (a) proton and (b) neutron $N=4$ and 5 major shells
to the angular momentum alignment $\langle J_x\rangle$ for
$\nu 1/2^-[550]$ in $^{101}$Pd.
The contributions of diagonal $\sum_{\mu} j_x(\mu)$ and off-diagonal part
$\sum_{\mu<\nu} j_x(\mu\nu)$ in Eq.~(\protect\ref{eq:j1})
from the proton and neutron $N=4$ shell are also shown as dashed lines.
}
\end{figure}

Contributions of proton and neutron $N=4$ and 5 major shells
to the angular momentum alignment $\langle J_x\rangle$ for
$\nu 1/2^-[550]$ in $^{101}$Pd are shown in Fig.~\ref{fig6:jxshell}.
The contributions of diagonal $\sum_{\mu} j_x(\mu)$ and off-diagonal part
$\sum_{\mu<\nu} j_x(\mu\nu)$ in Eq.~(\protect\ref{eq:j1})
from the proton and neutron $N=4$ major shell are also shown as dashed lines.
It can be clearly seen that the upbending for $\nu 1/2^-[550]$
in $^{101}$Pd at $\hbar\omega\sim$ 0.45~MeV
mainly comes from the contributions of the proton $N=4$ major shell,
while the neutron $N=4$ major shell does not contribute so much.

\begin{figure}[htbp]
\includegraphics[width=0.9\columnwidth]{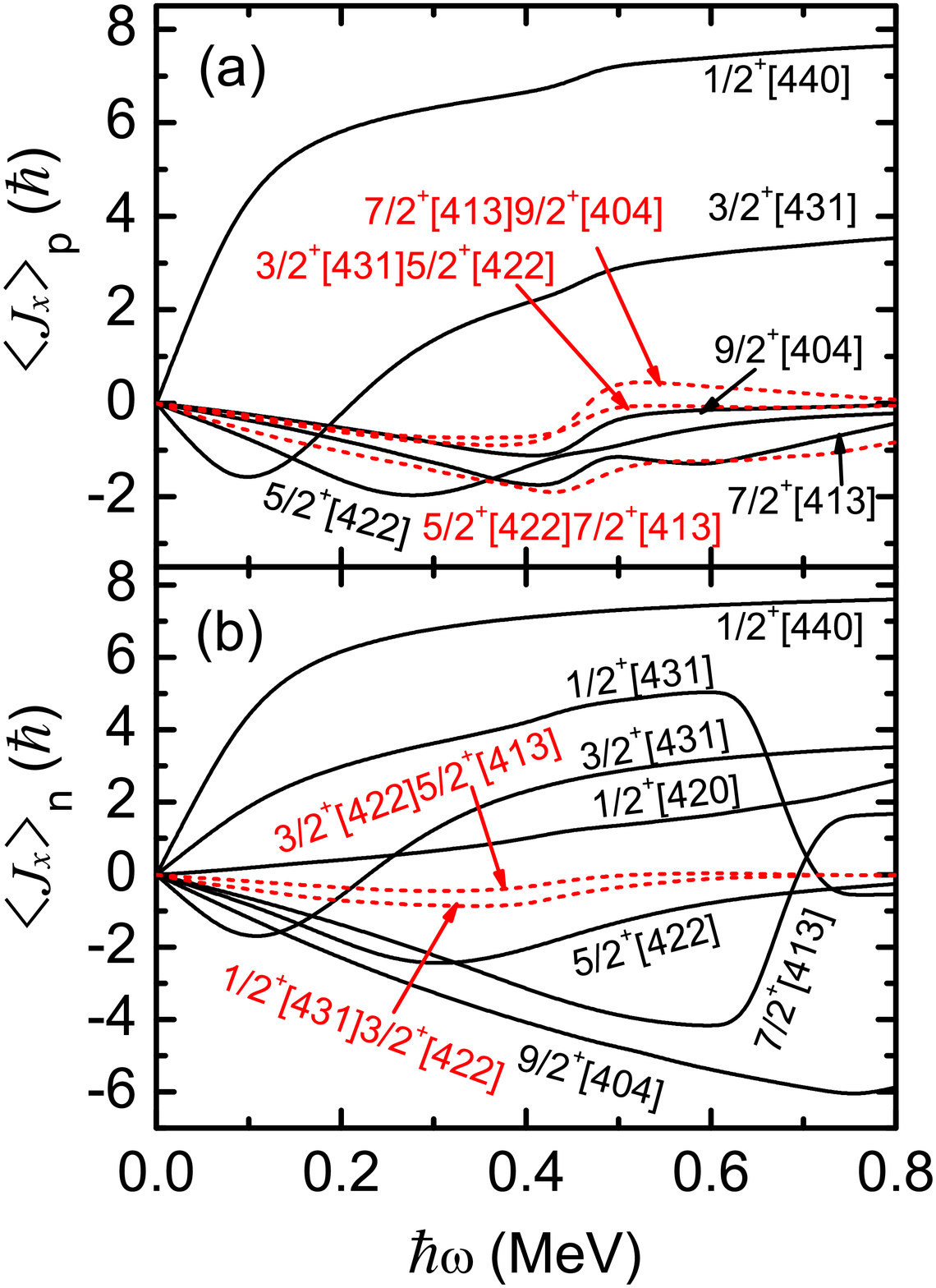}
\caption{\label{fig7:jxorb}(Color online)
Contributions of each (a) proton and (b) neutron orbital from
$N=4$ major shell to the angular momentum alignments
$\langle J_x\rangle$ for $\nu 1/2^-[550]$ in $^{101}$Pd.
The diagonal (off-diagonal) part $j_x(\mu)$
[$j_x(\mu\nu)$] in Eq.~(\protect\ref{eq:j1}) is denoted by black
solid (red dotted) lines.
}
\end{figure}

We note that for the proton $N=4$ major shell, both the diagonal and off-diagonal
parts contribute to the upbending, while for the neutron case,
only the off-diagonal part contributes.
This can be clearly seen from Fig.~\ref{fig7:jxorb},
where the contributions of each proton (top) and neutron (bottom)
orbital from $N=4$ major shell to the angular momentum alignments
$\langle J_x\rangle$ for $\nu 1/2^-[550]$ are presented.
The diagonal (off-diagonal) part $j_x(\mu)$
[$j_x(\mu\nu)$] in Eq.~(\protect\ref{eq:j1}) is denoted by black
solid (red dotted) lines.
In Fig.~\ref{fig7:jxorb}(a) for protons, one can easily find that
the diagonal parts
$j_x\left(\pi 7/2^+[413]\right)$ and $j_x\left(\pi 9/2^+[404]\right)$,
and the off-diagonal parts
$j_x\left(\pi 3/2^+[431] \pi 5/2^+[422]\right)$,
$j_x\left(\pi 5/2^+[422] \pi 7/2^+[413]\right)$, and
$j_x\left(\pi 7/2^+[413] \pi 9/2^+[404]\right)$
change a lot after upbending ($\hbar\omega\sim$ 0.45~MeV).
The alignment gain after the upbending mainly comes from these terms.
In Fig.~\ref{fig7:jxorb}(b) for neutrons,
one sees that the only contribution is from the off-diagonal part
$j_x\left(\nu 1/2^+[431] \nu 3/2^+[422]\right)$.
Again this demonstrates that the upbending  for $\nu 1/2^-[550]$
in $^{101}$Pd is mainly caused by the $\pi g_{9/2}$ orbitals,
and the contribution from the neutron $\nu g_{7/2}$
and $\nu d_{5/2}$ orbitals are rather small.

\begin{figure}[htbp]
\includegraphics[width=0.9\columnwidth]{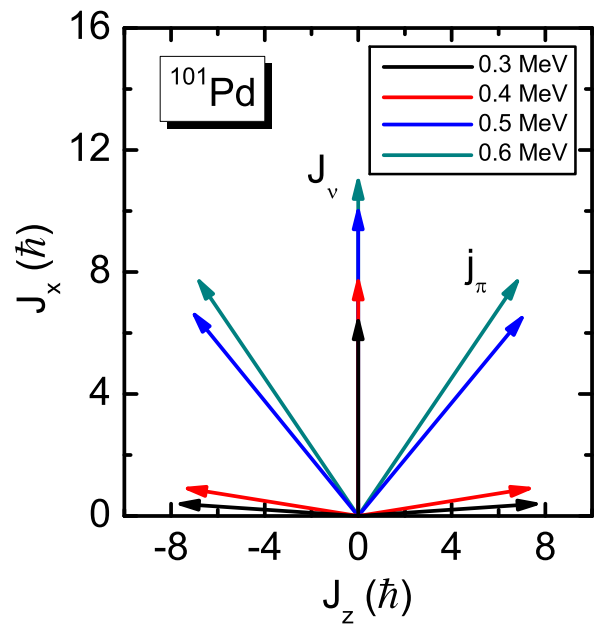}
\caption{\label{fig8:JxJz}(Color online)
Angular momentum vectors of neutrons
$J_\nu$ and the four $\pi g_{9/2}$ proton holes $j_\pi$,
at rotational frequencies from 0.3 to 0.6~MeV for
$\nu 1/2^-[550]$ in $^{101}$Pd.
Each proton angular momentum vector contains the contribution
of two $g_{9/2}$ proton holes.
}
\end{figure}

In order to examine the two-shears-like mechanism for
$\nu 1/2^-[550]$ in $^{101}$Pd,
the angular momentum vectors of neutrons
$J_\nu$ and the four $\pi g_{9/2}$ proton holes $j_\pi$ at rotational
frequencies from 0.3 to 0.6~MeV are shown in Fig.~\ref{fig8:JxJz}.
Each proton angular momentum vector contains the contribution of
two $g_{9/2}$ proton holes.
It should be noted that
the angular momenta of the four proton holes could,
in principle, be extracted exactly from the TAC calculation.
Here, $J_z$ is calculated approximately in the following way
according to Ref.~\cite{Frauendorf1996_ZPA356-263}
\begin{equation}
J_z = \sqrt{\langle \Psi | J_z^2 | \Psi \rangle} \ .
\end{equation}
This method has been proved to be a good approximation
by comparing the principal axis cranking with
the particle rotor model in Ref.~\cite{Frauendorf1996_ZPA356-263}
and has already been used for investigating the two-shears-like mechanism
in $^{105}$Cd and $^{106}$Cd by PNC-CSM~\cite{Zhang2013_PRC87-054314}.
It can be seen from Fig.~\ref{fig8:JxJz} that the two proton angular momentum
vectors $j_\pi$ are pointing opposite to each other and are nearly
perpendicular to the vector $J_\nu$ at $\hbar\omega = 0.3$~MeV.
The abrupt increasing of neutron angular momentum alignment
from $\hbar\omega =0.4$ to 0.5~MeV in Fig.~\ref{fig8:JxJz}
is due to alignment of the neutrons in $\nu g_{7/2}$ and $\nu d_{5/2}$ orbitals.
With increasing rotational frequency,
the higher angular momentum is generated by gradually closing
of the two blades of the proton angular momentum $j_\pi$
toward the neutron angular momentum vector $J_\nu$,
while the direction of the total angular momentum stays unchanged.
This leads to the closing of the two shears.
The two-shears-like mechanism is, thus, clearly seen.
It should be noted that from $\hbar\omega =0.4$ to 0.5~MeV,
the two shears close rapidly with increasing rotational frequency,
which is caused by the rearrangement of the proton occupations in $\pi g_{9/2}$ orbitals,
and the magnitude of two proton angular momentum vectors keep no longer constant.
This reflects the important role played by the proton angular momentum
alignment in the present two-shears-like mechanism in $^{101}$Pd.

\section{\label{Sec:Summary}Summary}

In summary, the possible antimagnetic rotation band $\nu 1/2^-[550]$
in $^{101}$Pd is investigated by the cranked shell model with
pairing correlations treated by a particle-number conserving method,
in which the blocking effects are taken into account exactly.
The experimental moments of inertia and reduced $B(E2)$ transition probabilities
are well reproduced by the PNC-CSM calculations.
By analyzing the $\omega$-dependence of the occupation probability of each
cranked Nilsson orbital near the Fermi surface and the contributions
of valence orbitals in each major shell to the total angular momentum alignment,
the upbending mechanism of $\nu 1/2^-[550]$ in $^{101}$Pd is understood clearly.
The upbending around $\hbar\omega \sim 0.45$~MeV is mainly caused by
the rearrangement of proton occupations in $g_{9/2}$ orbitals,
while the contribution from the neutron $g_{7/2}$ and $d_{5/2}$ orbitals is rather small.
Moreover, it is found that the proton angular momentum alignment,
which mainly comes from the rearrangement of proton occupations in $g_{9/2}$ orbitals,
plays also an important role in the two-shears-like mechanism.

\begin{acknowledgements}

Helpful discussions with Q. B. Chen, J. Peng, S. Q. Zhang, P. W. Zhao
and S. G. Zhou are gratefully acknowledged.
This work was partly supported by the Fundamental Research Funds
for the Central Universities (2015QN21), and the National Natural Science
Foundation of China (Grants No. 11275098, 11275248, 11505058).
The computational results presented in this work have been obtained
on the High-performance Computing Cluster of SKLTP/ITP-CAS and
the ScGrid of the Supercomputing Center, Computer Network Information Center
of the Chinese Academy of Sciences.

\end{acknowledgements}


%

\end{document}